\documentclass{aa}
\usepackage[varg]{txfonts}

\usepackage{graphicx}
\usepackage{textcomp}
\usepackage{color}
\usepackage{bbold}
\usepackage{breqn}
\usepackage{amsmath}
\usepackage{pstricks} 
\usepackage{bm} 
\usepackage{relsize}
\usepackage{footmisc}
\usepackage{tabularx}
\usepackage{array}
\usepackage[normalem]{ulem}

\usepackage[colorlinks=true,urlcolor=blue,citecolor=blue,linkcolor=blue,breaklinks=true]{hyperref}
\usepackage{twoopt}

\def\xlinkspace#1 #2{%
 \ifx\relax#2%
 \xlinkdash#1-\relax
 \else
 \xlinkdash#1 -\relax
 \expandafter\xlinkspace\expandafter#2%
 \fi}

\def\xlinkdash#1-#2{%
 \ifx\relax#2%
 \tmp{#1}%
 \else
 \tmp{#1-}%
 \expandafter\xlinkdash\expandafter#2%
 \fi}

\bibpunct{(}{)}{;}{a}{}{,} 

\makeatletter

 \newcommandtwoopt{\citeads}[3][][]{%
   \nonstopmode
   \href{http://adsabs.harvard.edu/abs/#3}%
        {\def\hyper@linkstart##1##2{}%
         \let\hyper@linkend\@empty\citealp[#1][#2]{#3}}
   \biblink{#3}{\href{http://adsabs.harvard.edu/abs/#3}{ADS}}%
   \errorstopmode}            
   
 \newcommandtwoopt{\citepads}[3][][]{%
   \nonstopmode
   \href{http://adsabs.harvard.edu/abs/#3}%
        {\def\hyper@linkstart##1##2{}%
         \let\hyper@linkend\@empty\citep[#1][#2]{#3}}
   \biblink{#3}{\href{http://adsabs.harvard.edu/abs/#3}{ADS}}
   \errorstopmode}            
   
 \newcommandtwoopt{\citetads}[3][][]{%
   \nonstopmode
   \href{http://adsabs.harvard.edu/abs/#3}
        {\def\hyper@linkstart##1##2{}%
         \let\hyper@linkend\@empty\citet[#1][#2]{#3}}
   \biblink{#3}{\href{http://adsabs.harvard.edu/abs/#3}{ADS}}%
   \errorstopmode}            
   
 \newcommandtwoopt{\citeyearads}[3][][]{%
   \nonstopmode
   \href{http://adsabs.harvard.edu/abs/#3}%
        {\def\hyper@linkstart##1##2{}%
         \let\hyper@linkend\@empty\citeyear[#1][#2]{#3}}
   \biblink{#3}{\href{http://adsabs.harvard.edu/abs/#3}{ADS}}%
   \errorstopmode}            
\makeatother

\makeatletter
\newcommand{\bibnote}[2]{\@namedef{#1note}{#2}}
\newcommand{\biblink}[2]{\@namedef{#1link}{#2}}
\makeatother

\newcommand{\HeI}{\ion{He}{i}}
\newcommand{\HeII}{\ion{He}{ii}}
\newcommand{\HeIII}{\ion{He}{iii}}
\newcommand{\CaII}{\ion{Ca}{ii}}
\newcommand{\MgII}{\ion{Mg}{ii}}
\newcommand{\CaIIK}{\CaII~K}
\def\CaIIHK{\CaII~H\&K}
\newcommand{\MgIIhk}{\ion{Mg}{ii}~h\&k}
\newcommand{\SiIV}{\ion{Si}{IV}}
\def\HI{\ion{H}{I}}

\def\Halpha{\mbox{H$\alpha$}}
\def\kms{\mbox{km s$^{-1}$}}

\newcommand{\be}{\begin{equation}}
\newcommand{\ee}{\end{equation}}

\newcommand{\bea}{\begin{eqnarray}}
\newcommand{\eea}{\end{eqnarray}}
\newcommand{\Heline}{\ion{He}{i}~1083~nm}
\newcommand{\Caline}{\ion{Ca}{ii}~854.2~nm}
\newcommand{\Feline}{\ion{Fe}{i}~617.3~nm}
\newcommand{\Siline}{\ion{Si}{i}~1082.7~nm}
\newcommand{\va}{\ensuremath{v_\mathrm{A}}}

\begin{document}

\title{High flow speeds and transition-region like temperatures in the solar chromosphere during flux emergence}
\subtitle{Evidence from imaging spectropolarimetry in \Heline\ and numerical simulations}
\titlerunning{High flow speeds and transition-region like temperatures in the solar chromosphere}
\authorrunning{Leenaarts et al.}


\author{
  J.~Leenaarts\inst{1} \and
  M.~van Noort\inst{2} \and
  J.~de la Cruz Rodr\'{i}guez\inst{1} \and
  S.~Danilovic\inst{1} \and
  C.~J.~D\'{i}az Baso\inst{3,4} \and
  T.~Hillberg\inst{1} \and
  P.~S{\"u}tterlin\inst{1} \and
  D.~Kiselman\inst{1} \and
  G.~Scharmer\inst{1} \and
  S.~Solanki\inst{2}}

\offprints{J. Leenaarts \email{jorrit.leenaarts@astro.su.se}}

\institute{Institute for Solar Physics, Dept. of Astronomy, Stockholm University, AlbaNova University Centre, SE-106 91 Stockholm, Sweden
\and
Max-Planck-Institut f{\"u}r Sonnensystemforschung, Justus-von-Liebig-Weg 3, 37077 G{\"o}ttingen, Germany
\and
Institute of Theoretical Astrophysics, University of Oslo, PO Box 1029, Blindern 0315, Oslo, Norway
\and 
Rosseland Centre for Solar Physics, University of Oslo, PO Box 1029, Blindern 0315, Oslo, Norway
}

\date{Received; Accepted }

\abstract 
{Flux emergence in the solar atmosphere is a complex process that causes a release of magnetic energy as heat and acceleration of solar plasma at a variety of spatial scales.}
{We aim to investigate temperatures and velocities in small-scale reconnection episodes during flux emergence.}
{We analysed imaging spectropolarimetric data taken in the \Heline\ line with a spatial resolution of $0.26\arcsec$,
a time cadence of 2.8\,s, and a spectral range corresponding to 180~\kms\ around the line. This line is sensitive to temperatures larger than 15~kK, unlike diagnostics such as \MgIIhk, \CaIIHK, and \Halpha, which lose sensitivity already at 15~kK. The \HeI\ data is complemented by imaging spectropolarimetry in the \CaIIK, \Feline, and \Caline\ lines at a cadence between 12\,s and 36\,s. We employed inversions to determine the magnetic field and vertical velocity in the solar atmosphere. We computed \Heline\ profiles from a radiation-MHD simulation of the solar atmosphere to help interpretation of the observations.}
{We find fast-evolving blob-like emission features in the \Heline\ triplet at locations where the magnetic field is rapidly changing direction, and these are likely sites of magnetic reconnection. We fit the line with a model consisting of an emitting layer located below a cold layer representing the fibril canopy. The modelling provides evidence that this model, while simple, catches the essential characteristics of the line formation. The morphology of the emission in the \Heline\ is localized and blob-like, unlike the emission in the \CaIIK\ line, which is more filamentary.}
{The modelling shows that the \Heline\ emission features and their Doppler shifts can be caused by opposite-polarity reconnection and/or horizontal current sheets below the canopy layer in the chromosphere. 
Based on the high observed Doppler width and the blob-like appearance of the emission features, we conjecture that at least a fraction of them are produced by plasmoids.
We conclude that transition-region-like temperatures in the deeper layers of the active region chromosphere are more common than previously thought. 
}

    \maketitle
\section{Introduction} \label{sec:intro}

Flux emergence in the solar atmosphere is a complex process that causes a release of magnetic energy as heat and acceleration of solar plasma at a variety of spatial scales.
In solar active regions, magnetic flux emerges from the convection zone into the photosphere at granular scales. The associated magnetic flux ropes have an undulating character dipping in to and out of the convection zone
\citep{2004ApJ...614.1099P}.
The emerging magnetic field reconnects with itself at progressively larger scales and interacts with previously emerged fields until longer loops are formed that can span from one polarity in the active region to the other
\citep[e.g][]{2016ApJ...825...93O}.
The flux emergence is associated with heating of the chromosphere as evidenced by the increased brightness and radiative losses in active regions compared to quiet Sun
\citep[e.g.][]{1977ARA&A..15..363W}

Flux emergence is associated with various brightening events and acceleration of plasma on a variety of scales
\citep[e.g.][]{2024A&A...686A.218N}.
On sub-granular scales, self-reconnection in the photosphere gives rise to \Halpha\ Ellerman bombs
\citep[e.g.][]{Georgoulis_2002,2011ApJ...736...71W};
at larger heights this process might appear as a UV-burst in observations of the \SiIV\ 140~nm line
\citep[e.g.][]{2014Sci...346C.315P,2015ApJ...812...11V,2019A&A...627A.101V}.
On scales of the order of a few granules, reconnection and electric currents give rise to thread-like brightenings in \CaIIHK\ and \CaII\ infrared triplet lines as well as a more diffuse brightening surrounding the flux emergence sites
\citep{2015ApJ...810..145D,2018A&A...612A..28L}.
On scales exceeding 10 arcseconds, chromospheric fibrils can become bright all along their length
\citep[][]{2016ApJ...825...93O}.

Currents sheets associated with magnetic reconnection are susceptible to the tearing instability and tend to fragment, forming magnetic islands or plasmoids in 2D geometry, and a complex structure of twisted magnetic flux ropes in 3D geometry
\citep[see Sec. 8 of][and references therein]{2022LRSP...19....1P}.
In simulations, these plasmoids have typical size ranging from order one to a hundred kilometers, at the limit or below the diffraction limit of meter-class telescope. They have complex internal structure, with plasma at chromospheric and transition region temperatures lying close together and substantial velocity gradients. In 2D geometry, they are expelled from the reconnection site with speeds of the order of the Alfv{\'e}n velocity. In observations, plasmoids appear as small bright blobs in the \CaIIHK\ and \Halpha\ lines
\citep{2017ApJ...851L...6R,2023A&A...673A..11R,2021A&A...647A.188D}.

Ellerman bombs and UV bursts and their observational signature have been modeled using 3D radiation-MHD simulations
\citep{2017ApJS..229....5D,2017A&A...601A.122D,2017ApJ...839...22H,2019A&A...626A..33H}.
Current sheets did not fragment into plasmoids and small-scale twisted flux ropes in these simulations because of the limited resolution in such 3D simulations.

Resolution in 2D simulations is much less of a concern, and plasmoid formation in 2D under  chromospheric and transition-region-like circumstances has been simulated by various authors
\citep[e.g.][]{2015ApJ...799...79N,2017ApJ...851L...6R,2020ApJ...901..148G}.
\citet{2024A&A...685A...2C} performed 2D simulations with adaptive mesh refinement of flux emergence with a resolution of $\sim 1$~km. The simulation exhibited turbulent reconnection and plasmoid formation in the chromosphere away from photospheric flux cancellation sites. It is more reminiscent of reconnection when emerging short loops are pushed into pre-existing longer loops.

The high-temperature gas present during reconnection cannot be seen with \MgIIhk, \CaIIHK, and \Halpha\ because their ionisation stages have too low population at temperatures above 15~kK. Spectral lines in the UV offer this sensitivity but can only be observed from space. Slit spectroscopy of the \SiIV\ 140~nm lines with the Interface Region Imaging Spectrograph 
\citep[IRIS;][]{2014SoPh..289.2733D}
has been very successful at a resolution of $0.35\arcsec$, but requires scanning to obtain images, lowering the time resolution.
 
The atomic structure of \HeI\ makes the \Heline\ line sensitive to temperatures above 15~kK, and is observable from the ground, making it a unique diagnostic for studying heating in the chromosphere. The reason for the sensitivity at high temperatures is that the triplet system of \HeI\ is only significantly populated through either recombination from \HeII\ if illuminated by UV radiation 
\citepads{1939ApJ....89..673G}, 
or bombardment by a non-thermal electron beam
\citep{2005A&A...432..699D,2021ApJ...912..153K},
or through direct collisional excitation
\citepads{1973ApJ...186.1043M}.
The latter mechanism requires temperatures in the line-forming region higher than 20~kK, the former two do not.

The line is normally in absorption, and the spatial variation of the line strength is mainly caused by the variation of the UV flux impinging on the chromosphere
\citepads{2016A&A...594A.104L}.
However, the line can show emission in EB's
\citep{2017A&A...598A..33L}
and flares
\citep[e.g.][]{2014ApJ...793...87Z}.
\citetads{2021A&A...652A.146L} 
investigated the formation of the \Heline\ line in the simulations of a reconnection-driven bidirectional jet of
\citet{2019A&A...626A..33H}.
They self-consistently included UV radiation produced by hot regions in the simulation, but did not include the effect of non-thermal electrons. They found \Heline\ line profiles in emission relative to the continuum intensity. Population in the triplet system was driven by UV radiation produced by a pocket of material at coronal temperatures in the chromosphere, while the emission itself was caused by a shell of material of chromospheric temperature around this hot pocket of sufficient optical thickness to cause the source function to thermalise.


Direct collisional excitation of the triplet system in 3D MHD simulations has to our knowledge not been reported, but there is no a priori reason why this would not occur given sufficient material at the right combination of temperatures and densities. However, as argued by 
\citetads{2014ApJ...793...87Z},
there is only a limited range of temperatures where this excitation can populate the triplet system without also collisionally ionising most \HeI\ to \HeII. Non-equilibrium ionisation could sustain \HeI\ at temperatures higher than where it would be ionised away in equilibrium circumstances, but only for a short time (ranging from $\sim 100$~s at 23 kK, to 0.05~s at 93 kK). On the other hand, the optical thickness of the emitting layer does not need to be large to create emission relative to the continuum, because both the photon destruction probability $\epsilon$ and the Planck function are steeply increasing functions of temperature. 

Collisional ionisation by non-thermal particles would act the same as photo-ionisation, but requires non-thermal particles inside small-scale reconnection events. To our knowledge, no observational evidence for this has been reported. 

Here we use the first science-ready integral-field spectropolarimetry data in the \Heline\ line obtained with the Helium Spectropolarimeter (HeSP) instrument at the Swedish 1-m Solar Telescope 
%
%
to study small-scale reconnection events. We analyse the data using inversions and use radiation-MHD simulations of the solar atmosphere combined with non-LTE radiative transfer calculations of the \Heline\ to interpret the observations. 

\section{Observations and data reduction}\label{sec:observations}

We use observations from the Swedish 1-m Solar Telescope 
\citep{2003SPIE.4853..341S}
taken on 2023-06-29 from 08:48:41UT to 09:01:43UT.

The CRISP imaging spectropolarimeter
\citepads{2006A&A...447.1111S,2008ApJ...689L..69S,2021AJ....161...89D}
observed the \Feline\ line, \Halpha, and the \Caline\ with a cadence of 36~s. The \Feline\ line was sampled with full-Stokes polarimetry at 11 wavelength positions equidistantly spaced from -17.5~pm to +17.5~pm around line center and a single point in the nearby continuum at +42~pm from line center. \Halpha\ was sampled at 19 equidistant positions  from -90~pm  to +90~pm around line center with spectroscopy only. The  \Caline\ was sampled equidistantly with 15 points from -67.5~pm to +67.5~pm around line center with full-Stokes polarimetry. The pixel scale was 0.044 arcsec. 

The CHROMIS imaging spectrograph 
\citepads{2006A&A...447.1111S,2017psio.confE..85S}
sampled the \CaIIK\ line at 4 points placed at $\pm$150~pm, $\pm$117~pm, and at 25 points spaced equidistantly between -91~pm and +91~pm around line center. A point in the nearby continuum at 400~nm was observed as well. The cadence of the CHROMIS observations was 11.8\,s, and the pixel scale 0.038 arcsec. The CRISP and CHROMIS data were reduced with the SSTRED pipeline
\citepads{2015A&A...573A..40D,2021A&A...653A..68L}.
The polarimetric calibration of the CRISP data was performed by using a field-dependent demodulation matrix (as proposed by 
\citeads{2008A&A...489..429V}).
%

\begin{figure}
\includegraphics[width=\columnwidth]{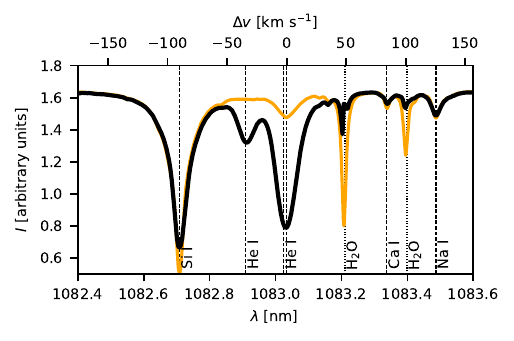}
\caption{Average spectrum and most important spectral lines in the HeSP observations. Black: spatially and temporally averaged HeSP observations. Orange: Atlas profile from \citet{1984SoPh...90..205N}, arbitrarily scaled to match the continuum of the HeSP data. Solar spectral lines are indicated with dashed vertical lines and telluric water lines with dotted lines.}
\label{fig:HeSP_avg_spectrum}
\end{figure}

\begin{figure*}
\centering
\includegraphics[width=\textwidth]{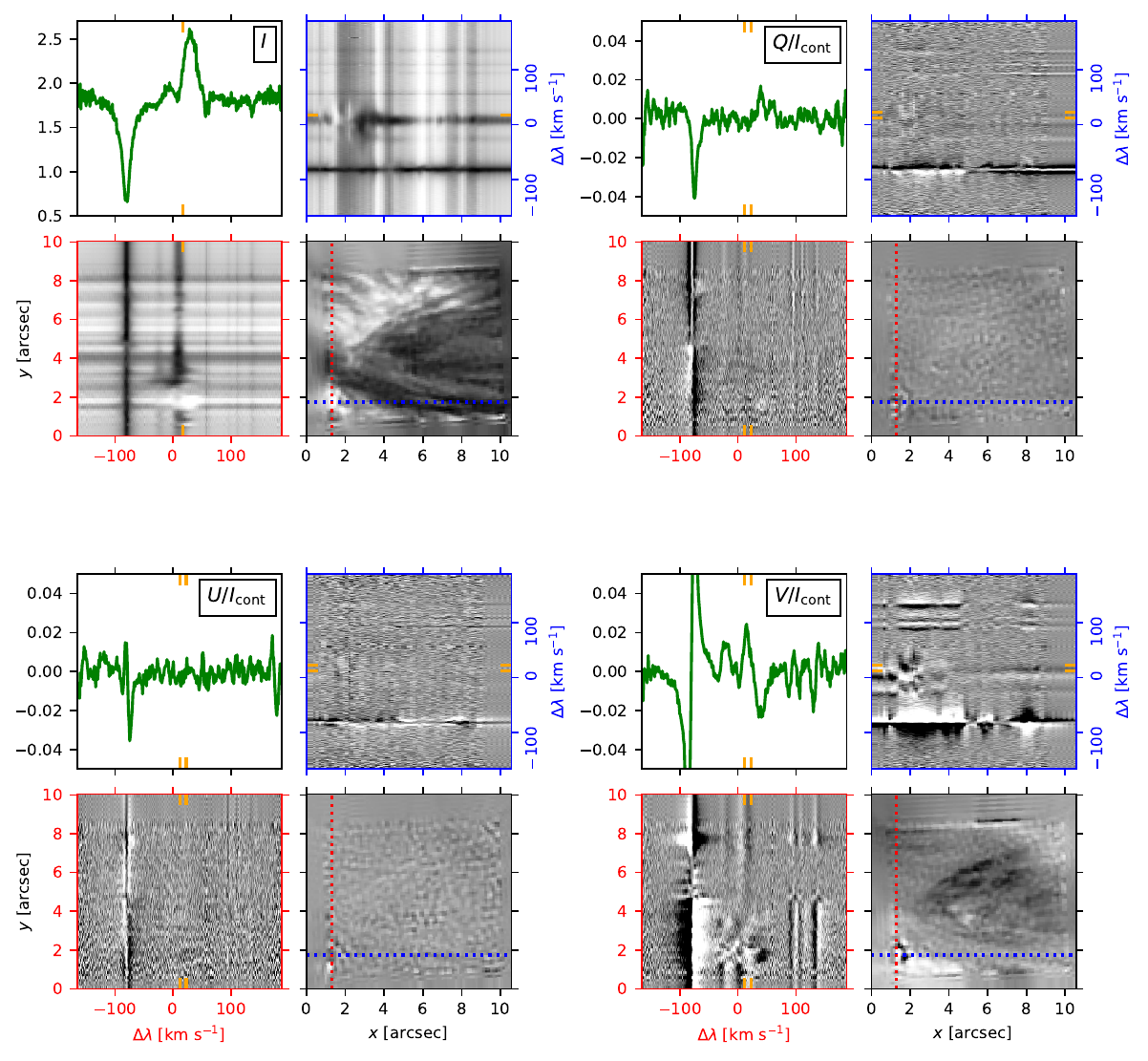}
\caption{Example HeSP snapshot, at 31~s after start of the observations. The four groups of four panels display Stokes $I$ (upper left), $Q$ (upper right), $U$ (lower left), and $V$ (lower right). Each group consists of four subpanels, showing an example line profile (upper left), an $x \lambda$-slice (upper right), a $\lambda y$-slice (lower left) and an $xy$-image (lower right). The locations of the $x \lambda$ and $\lambda y$-slices are indicated with blue and red dotted lines in the image, and their intersection is the location for which the line profile is shown. The wavelength of the Stokes $I$ image is indicated with the orange tick marks in the Stokes $I$ frames.  The $Q$, $U$, and $V$ images are wavelength averages, with the average taken over the interval indicated by the orange ticks in the respective frames. The Stokes $I$ intensity is given in arbitrary units. The wavelength axes are specified in units of Doppler velocity relative to $\lambda=1083.0$~nm. An animated version of the figure is available online.}
\label{fig:HeSP_overview}
\end{figure*}

A recent addition to the instrumentation, the Helium Spectropolarimeter (HeSP)
%
%
observed the \Heline\ line and the nearby \Siline\ line with full-Stokes polarimetry. The HeSP is a microlensed hyperspectral imager \citep[MiHI][]{2022A&A...668A.149V}, specifically optimized for the helium line at 1083~nm, with a field-of-view (FOV) of $59\times67$ spatial pixels, each with an angular size of 0.13 arcsec, which approximately corresponds to critical sampling according to the Rayleigh criterion at this wavelength, 
yielding a FOV of $7.7\arcsec \times 8.7$\arcsec.
The spectral domain is sampled with 4~pm pixels from 1082.2~nm to 1083.8~nm corresponding to a Doppler speed of about $\pm$220~\kms\ around the \Heline\ line. 
The raw HeSP data were reduced with a procedure closely following the one used for the MiHI prototype
\citepads{2022A&A...668A.149V,2022A&A...668A.150V,2022A&A...668A.151V}.
%
%
%

The HeSP cameras operate at a framerate of 113.6~Hz. In sit-and-stare mode, as in the observations used here, the cadence of the reduced data can be decided after the observations. Here we reduced the data with a cadence of 2.8~s. Based on the root mean square (RMS) variations of the noise in the continuum on the red side of the lines, we estimate the noise floor to be about $1\times10^{-2}$ for a single $(x,y,\lambda)$ pixel.

The target was the active region NOAA 13354. CRISP, CHROMIS, and HeSP have overlapping fields-of-view (FOVs) and observed cotemporally. The small HeSP FOV was approximately centred on $(x,y)=(80\arcsec,200\arcsec)$ in helioprojective Cartesian coordinates (corresponding to $\mu = 0.98$), more or less in the middle of the FOV of the other two instruments.

The CRISP, CHROMIS, and HeSP data were coaligned to within one CHROMIS pixel, and resampled in space to the CHROMIS pixel scale of $0.0379\arcsec$. The resampling of the HeSP data was done using nearest neighbour interpolation, for CRISP data linear interpolation was used. We chose to use nearest neighbour interpolation for HeSP because the large difference in spatial resolution between the \Heline\ and \CaIIK\ would otherwise lead to artificially smooth HeSP images. 
The CRISP and CHROMIS data was resampled in time to the HeSP cadence of 2.8~s using nearest-neighbour interpolation, again to avoid artificial smoothness, but now in the time dimension. 
We chose not to rotate the data such that solar north points up to avoid unnecessary smoothing of the data. Therefore, the origin and orientation of the $x$ and $y$ axes in the figures are arbitrary.

Figure~\ref{fig:HeSP_overview} shows an example frame of the HeSP data. The images appear sharp and at diffraction-limited resolution,
and the spectra show strong variation on sub-arcsecond scales. The \Heline\ line is mostly in absorption, but also exhibits strong emission at certain locations in the FOV. 

Because of the limited signal-to-noise ratio per pixel, Stokes $Q$ and $U$ are far above the noise in the \Siline\ only, but the photospheric \ion{Ca}{i} and \ion{Na}{i} lines (see Fig.~\ref{fig:HeSP_avg_spectrum}) also show a weak signal. The removal of the telluric water 
lines introduces weak spurious $Q$ and $U$ signal. In \Heline,  $Q$ and $U$ are clearly above the noise only at a few locations, such as the $Q$ vs $\lambda$ plot and the $x \lambda$-cut in the upper right of the figure. In contrast, Stokes $V$ is clearly visible in \Heline\ in many locations in the image and the spectra.

\section{Inversions} \label{sec:inversions}

\subsection{Inversions of the \Feline\ and \Caline\ lines}
The \Feline\ data was used in a spatially-regularized Milne-Eddington inversion following
\citetads{2019A&A...631A.153D}
to estimate the photospheric magnetic field vector.

We obtained an estimate of the chromospheric magnetic field vector by applying a spatially-regularized weak-field approximation method to the \Caline\ data following 
\citetads{2020A&A...642A.210M}.
We did not attempt to resolve the 180-degree ambiguity in the plane-of-the-sky (POS) in either of these inversions.

\begin{figure*}
\centering
\includegraphics[width=\textwidth]{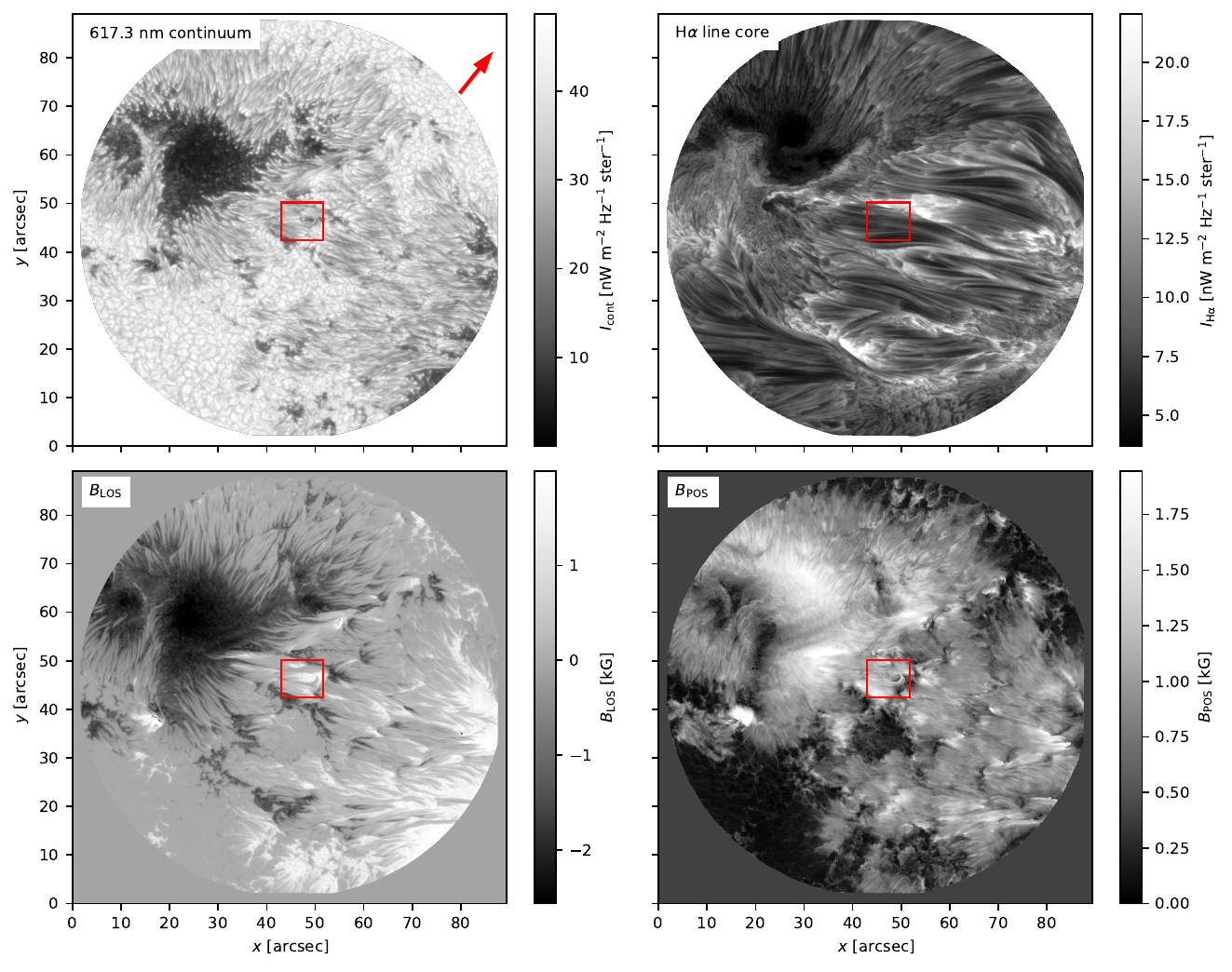}
\caption{Overview of the observed region at 2023-06-29T08:49, the start of the observation. The upper row shows the continuum close to the \Feline\ line and the \Halpha\ line core. The bottom row shows the line-of sight and plane-of-the-sky components of the photospheric magnetic field, as determined from the spatially-coupled Milne-Eddington inversion of the \Feline. 
The red box shows the HeSP FOV and the arrow in the upper left panel points toward solar North.}
\label{fig:FOV_overview}
\end{figure*}

\subsection{Inversion of the \Heline\ line}\label{sec:model_fits}

Selected pixels in the \Heline\ data were inverted using the {\sc Hazel-2} code
\citepads{2008ApJ...683..542A},
which can treat the full quantum formation of the He polarised signals including the Hanle, Zeeman and Paschen-Back effects. {\sc Hazel-2} solves the radiative transfer in a constant-property slab model using a five-term atomic model for triplet state helium taking into account only radiative transitions. The emergent intensity is determined via an analytical solution to the polarized radiative transfer equation given by:
\begin{equation}
{\bf I}={\rm e}^{-{\mathbf{K}^{*}}\Delta\tau}\,{\bf I}_{0}\,+\,\left[{\mathbf{K}^{*}}\right]^{-1}\,
\left( \mathbf{1} - {\rm e}^{-{\mathbf{K}^{*}}\Delta\tau} \right)\,\beta \mathbf{S}. \label{eqn:slab_model}
\end{equation}
$\bf I_{0}$ is the Stokes vector incident on the boundary of the constant-property slab, furthest from the observer, and along the same line-of-sight. $\mathbf{K}^{*}$ is the propagation matrix normalized by its first element ($\eta_{I}$), $\Delta\tau$ is the optical thickness of the slab, and $\mathbf{S}$ is the source function vector. The quantity $\beta$ is an ad hoc scalar parameter that scales the source function to account for deviations from the collision-free case. We treat it as a free parameter. In the optically thin case $\beta$ is ill-constrained and better kept fixed at unity, but for the pixels that we invert this is not the case.

The version of {\sc Hazel-2} that we employ inverts the photospheric \Siline\ line to provide an estimate of the incoming intensity ${\bf I}_{0}$ for the first slab above the photosphere. This photospheric inversion is done with the {Stokes Inversion based
on Response functions} code \citep[SIR;][]{1992ApJ...398..375R}, that assumes LTE and hydrostatic equilibrium to solve the radiative transfer equation.

Here we focus the inversions on those pixels which show emission in the \Heline\ line. The observations show a dark layer of fibrils in the core of \Heline\ covering most of the FOV, and giving the impression that these fibrils are located higher in the atmosphere than most sites that produce the emission (see also Sec.~\ref{subsec:zoomin} and Fig.~\ref{fig:all_images}). 

Radiative transfer modelling of an Ellerman bomb/UV burst by
\citetads{2021A&A...652A.146L} 
indicates that the contribution function of the \Heline\ is non-zero in only a few layers of only roughly 100~km extent in the vertical direction. 

Taken together this led us to use {\sc Hazel-2} with a stacked two-slab model, one located deeper down in the atmosphere, which we assume to be the emission site, and one slab higher in the atmosphere that represents the dense canopy of fibrils. The intensity from the photosphere is fed into the emission site slab, and the intensity from this slab is used as a boundary for the fibril slab. A similar approach was used in 
\citet{2017A&A...598A..33L}.

Owing to the noise level (of $10^{-2}$ in units of continuum intensity) and weak linear polarization signals in our FOV, we will not 
be able to infer the horizontal component of the magnetic field, so these parameters are kept at zero Gauss.

In order to characterize the incident radiation field, we need to specify the heights of our slabs above the photosphere. In our case, these cannot be inferred since we observe the region from above. We assumed a height of 2.2 Mm for the emission site slab as well as for the fibril canopy slab. While using the same height is unphysical, we cannot determine the heights of the slabs in our observations, so we stick with a single value for simplicity. The polarization signals are relatively insensitive to the exact values of the heights, and might lead to variations in the inferred field on the order of 10~G, so this is not important for our results
\citep{2019A&A...625A.128D,2019A&A...625A.129D}. 

In total there are sixteen 
free parameters per pixel: we fit three nodes in temperature and one in line-of-sight velocity in the photospheric inversion. In each of the two slabs we fit the line-of-sight magnetic field $B_{\rm LOS}$, the optical thickness at the center of the red blended component of the \Heline\ multiplet $\Delta\tau$, the line-of-sight velocity $v_{\rm LOS}$, the source function scaling factor $\beta$, the line width $\Delta v_D$ and the damping parameter of the Voigt profile $a$. 

We use the FAL-C model 
\citepads{1993ApJ...406..319F}
as starting guess model for the photosphere. Some manual intervention was needed for the He slabs to converge to the best-fit solution. We achieve satisfactory fits using the strategy described above (see Sec.~\ref{sec:results}).

{\sc Hazel-2} does not solve the full non-LTE problem. Instead, the source function is parametrized as a factor $\beta$ times the first irreducible spherical tensor component of the radiation intensity $J_0^0$, which is equal to the angle-averaged intensity (neglecting the minor contribution from $J_0^2$):
\be
S_\nu = \beta J_0^0 =\beta \oint \frac{d \Omega}{4\pi} I(\nu,\mu)
\ee
The intensity $I(\nu,\mu)$ used here is computed from the continuum irradiation from the solar photosphere. It takes into account the assumed height of the slabs, the center-to-limb variation and the wavelength dependence of the solar continuum radiation field $I(\nu,\mu)$ as given in Allen's astrophysical quantities \citep{Cox2000asqu.book.....C}.

Because we assumed the height of our slabs to be constant, we can compute the excitation temperature 
\citepads{2003rtsa.book.....R}
as function of $\beta$:
\be
T_{\mathrm{exc}}(\nu) = \frac{h\nu}{k_B} \left( \ln\left(1 + \frac{2 h \nu^3}{c^2 \beta J_0^0}\right)\right)^{-1},
\ee
where $h$ is the Planck constant and $k_B$ is the Boltzmann constant. For $\beta=1$, we find $T_{\mathrm{exc}} = 4.6$~kK, for $\beta=10$ we obtain $T_{\mathrm{exc}} = 13$~kK.
We do not find solutions with $\beta<1$, and the excitation temperature can therefore be interpreted as a lower bound of the gas temperature in the slab assuming the \Heline\ source function is described well by a two-level source function:
\be
S_\nu = (1-\epsilon) J_0^0 + \epsilon B_\nu,
\ee
with $\epsilon$ the photon destruction probability. Because we cannot determine $\epsilon$, the excitation temperature is a lower bound for the gas temperature. 

\section{Numerical modelling}\label{sec:simulations}

To help interpret the observations, we employed a snapshot from a radiation-MHD simulation of the solar atmosphere, and solved the non-LTE radiative transfer problem for the \Heline\ line for this snapshot.   
The 3D model atmosphere was produced with the MURaM code  \citep{2005A&A...429..335V,2017ApJ...834...10R}. The simulation is the flux emergence run 1 from \citet{2023AdSpR..71.1939D}. It tries to reproduce flux emergence in a mildly active region of the Sun. A bipolar flux system is advected through the bottom boundary into a fully developed plage simulation that includes subsurface layers, down to 8 Mm, photosphere, chromosphere, and corona up to 14 Mm. The emergence is limited within an ellipsoidal area with a major and minor axis 
$(a,b) = (10,2)$~Mm and field strength of 5000 G \citep{2019NatAs...3..160C}. The emerging flux reaches the surface after approximately 5 hours of solar time. The grid spacing in the simulation domain is then reduced to half of the original value so that the final resolution is $39$ and $21$~km  in the horizontal and vertical direction, respectively. The simulation domain is $40 \times 40 \times 22$~Mm. 

We used a snapshot from the simulation as input atmosphere for the 3D non-TE radiative transfer computation performed with the Multi3d code
\citepads{2009ASPC..415...87L}.
The input electron density was computed assuming LTE.

Multi3d solves the statistical equilibrium equations using the formalism of 
\citetads{1992A&A...262..209R},
where the radiation field is evaluated in 3D using 24 angles in the A4 set of 
\citetads{carlson1963}.
Because the photoionisation-recombination mechanism is an essential mechanism that drives the population of the triplet system of \HeI, we included production of UV photons in the transition region and corona for wavelengths shortward of the \HeI\ ionisation edge at 50.4~nm
The coronal emissivity is computed as $j_\nu = \Lambda_\nu (T) n_{\mathrm{e}} n_{\mathrm{H}}$, where $ \Lambda_\nu(T)$ is computed using the CHIANTI package
\citepads{2009A&A...498..915D} assuming coronal equilibrium.
This emissivity is added to the background emissivity in Multi3d. A detailed description of this method is given in 
\citetads{2016A&A...594A.104L}.

We use the \ion{He}{}\ model atom described in 
\citetads{2021A&A...652A.146L}.
It consists of 16 energy levels, of which 12 levels are in \HeI, 3 are in \HeII, and the final one is the \HeIII\ continuum. The model includes the 1s\,2s\,$^3$S$_1$ and 1s\,2p\,$^3$P$_{0,1,2}$ levels that produce the three components of the \Heline\ line. A detailed description of the model atom including a term diagram is given in 
\citetads{2021A&A...652A.146L}.

In addition, we computed the emergent $\Halpha$ intensity with Multi3d using full 3D non-LTE radiative transfer following 
\citet{2012ApJ...749..136L}
and the \CaIIK\ intensity in non-LTE including the effects of PRD in the 1.5D approximation using the STiC code 
\citep{2019A&A...623A..74D,2001ApJ...557..389U}.

In addition to these new calculations, we also use a snapshot of a radiation-MHD simulation computed with the Bifrost code by
\citetads{2019A&A...626A..33H} to interpret our observations.
The radiative transfer problem for He in this snapshot was solved by 
\citetads{2021A&A...652A.146L}.
This snapshot contains an Ellerman-bomb /UV-burst small-scale reconnection event which shows emission in the \Heline\ line.

\section{Results}\label{sec:results}

\subsection{Overview of the observed region} 

Figure~\ref{fig:FOV_overview} shows an overview of the observed active region (AR) NOAA 13354 based on the CRISP data. The AR harbours two groups of sunspots of opposite polarity. One group is inside the CRISP FOV, the second group is outside the FOV at the lower right of the figure.  Inspection of SDO/HMI data of the AR shows that most flux had emerged by the time of the observations, but the sunspot groups continued to separate for a few days after this date, and flux emergence was still ongoing during the observations. This is evidenced by the elongated structures visible in the continuum intensity between the sunspot groups and the corresponding elongated photospheric magnetic field structures of both polarities visible in $B_{\mathrm{LOS}}$. The HeSP FOV is located in this area with flux emergence. The \Halpha\ image shows thick fibrils covering most of the FOV. The photospheric LOS magnetic field structure within the FOV is markedly different from, and predominantly of the opposite polarity as the field in the nearby penumbra.

\begin{figure*}
\centering
\includegraphics[width=\textwidth]{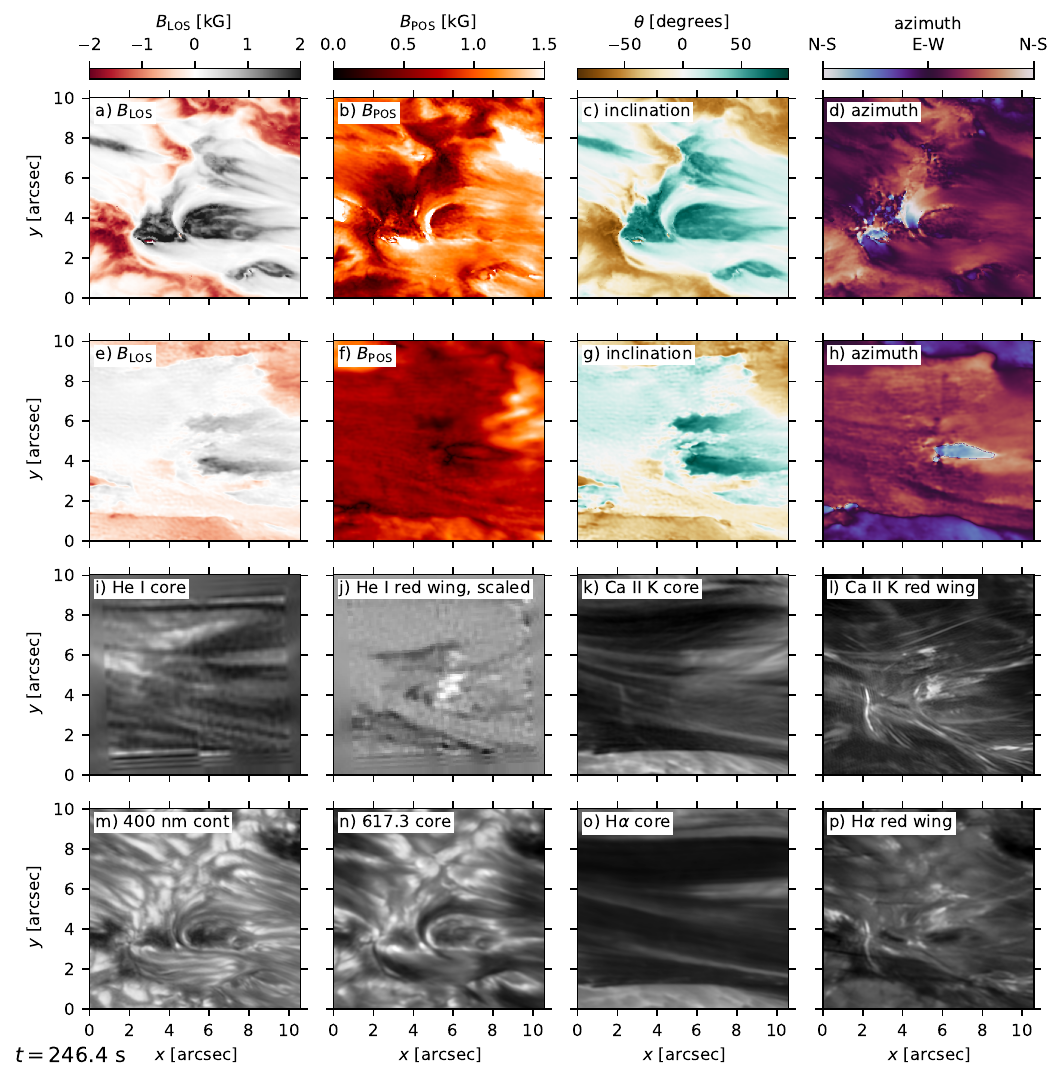}
\caption{Overview of the HeSP FOV at 246\,s after the start of the observations. First row (a-d): magnetic field in the photosphere as determined from a Milne-Eddington inversion of the \Feline\ recorded by CRISP. Second row: magnetic field in the chromosphere as determined from the weak-field approximation in the \Caline\ (also CRISP data). Bottom two rows: intensity images at various wavelengths. The H$\alpha$ red wing image is at $\Delta \lambda = +45$~\kms\ from the nominal line core, the \CaIIK\ red wing image at $\Delta \lambda = +36$~\kms.  Panel j) shows the intensity normalised to the local continuum intensity $I(x,y,\lambda)/I_{\mathrm{cont}}(x,y)$, 
the other panels show the unscaled intensity $I(x,y,\lambda)$. A movie of this figure showing the entire time series is available online. 
}
\label{fig:all_images}
\end{figure*}

\begin{figure*}
\centering
\includegraphics[width=\textwidth]{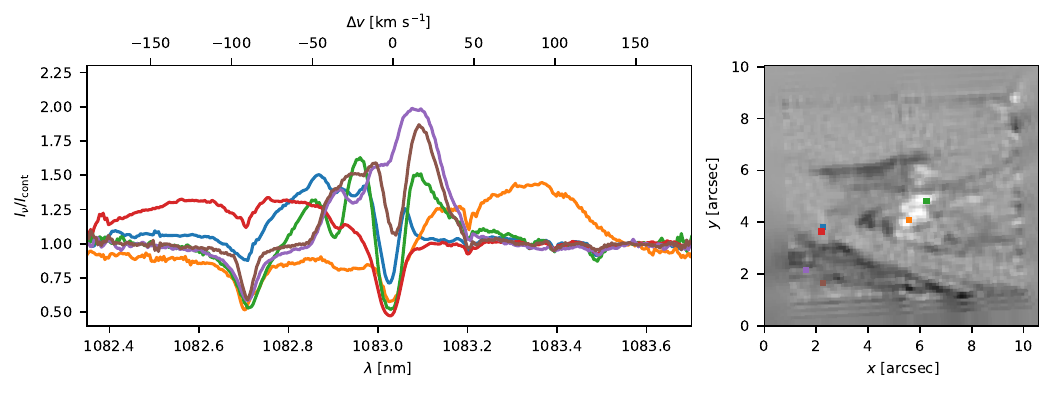}
\caption{Example \Heline\ line profiles that show emission. Left: line profiles. Right: local continuum normalised intensity in the red wing of \Heline, with the locations of the profiles left indicated with squares of the same colour as the profile. The line profiles are taken from the entire time series, not just from the timestep shown in the right-hand panel.}
\label{fig:emission_examples}
\end{figure*}

\subsection{Zoom in on the HeSP field-of-view} \label{subsec:zoomin}

Figure~\ref{fig:all_images} shows the HeSP FOV. It contains a number of small pores with different polarity (panels~{\it a} and~{\it m}). 

The photospheric magnetic field is rather vertical in the pores, but the FOV contains patches with kG-strength horizontal field as well (such as $(x,y) = (9",7")$), indicative of flux emergence. 
The azimuth of the magnetic field is predominantly NW to SE, except around the pores at $y=4"$, where the azimuth structure appears more complex.

The second row of the figure shows the chromospheric magnetic field as determined from the \Caline\ line. The vertical chromospheric field (panel~{\it e}) is much smoother than in the photosphere and has a lower unsigned field strength. Comparison with panels~{\it e} and~{\it o} shows that the positive LOS magnetic field is associated with the dark fibrils covering much of the FOV. Where the fibrils are not present the LOS magnetic field is negative. There are two elongated patches of strong vertical field at $y=4\arcsec$ and $y=6\arcsec$. The horizontal field and the inclination (panels~{\it f} and~{\it g}) show that in these patches the field is much more vertical. The azimuth panel shows that the horizontal direction of the field largely follows the orientation of the fibrils, except in a small area around $(x,y) = (7",4.5")$, where the azimuth is roughly 90 degrees different.

The picture emerging from this analysis is that there is ongoing small-scale flux emergence in the photosphere with opposite polarity fields existing close together and a complex structure of the horizontal direction of the field. The overlying upper chromosphere appears to consist mostly of ordered fibrils as seen in \Halpha\ and \CaIIK\ which are much longer than the extent of the HeSP FOV (see Fig.~\ref{fig:FOV_overview}). The atmosphere between the photosphere and the fibril canopy is therefore expected to show a complex evolving magnetic field exhibiting small-scale reconnection and current sheets with concomitant heating, flows and jets
\citepads[e.g.,][]{2020A&A...633A..58O,2021A&A...647A.188D}.

This view is corroborated in panels~{\it l} and~{\it p}, which show images at wavelengths in the red wing of the lines, chosen such that they show a scene in between the photosphere and the fibril canopy. The \CaIIK\ panel in particular is filled with elongated intermediate-brightness threads, which are sites with elevated temperature 
\citepads[e.g.][]{2018A&A...612A..28L}. It also shows blobs and shorter threads with high brightness.
The \Halpha\ panel shows fewer and less pronounced intermediate-brightness threads, but does show some high-brightness blobs and threads that partially coincide with the ones visible in \CaIIK. These phenomena are not visible in the core of the \Feline\ (panel {\it n}), indicating that the heating is occurring at least 100-200 km higher than the local formation height of the continuum
\citepads[e.g.][]{2023A&A...669A.144S}.

Finally, we turn to the appearance of the \Heline\ line. The figure does not show the 1083 nm continuum, but inspection shows it appears largely similar to the 400\,nm continuum in panel {\it m}, but with lower contrast and lower spatial resolution owing to the longer wavelength. The line core (panel {\it i}) shows fibrils with the same overall orientation as in the \Halpha\ and \CaIIK\ line cores. They appear thinner and not as dark, but are nevertheless completely optically thick, and no photospheric structure is visible. 
This is expected in active regions where the hot and dense corona produces copious amounts of UV photons to drive the photoionisation-recombination mechanism that populates the triplet system of \HeI.

Panel {\it j)} shows the intensity at $\lambda=1083.11$~nm, in the red wing of \Heline, divided by the local continuum intensity, i.e., $I(x,y,\lambda)/I_{\mathrm{cont}}(x,y)$. Dividing by the continuum makes the background appear flat, emission above the local continuum level appears bright and absorption features appear dark. The most prominent features are the emission structures at $(x,y)=(6\arcsec,4\arcsec)$. Inspection of the movie of Fig.~\ref{fig:all_images} shows that the emission appears as roundish blobs with a diameter smaller than $0\farcs5$ to elongated blobs of up to $2\arcsec$ length and a width below $0\farcs5$. Often, but not always there is emission in the red wings of \CaIIK, and \Halpha\ at locations of \Heline\ emission. The reverse is not true, many bright structures visible in \CaIIK, such as the threads at $(x,y)=(2\arcsec,3\arcsec)$ and $(x,y)=(9\arcsec,1.5\arcsec)$ do not have a counterpart in \Heline. The \Heline\ wing image also does not show the elongated intermediate brightness threads that are present in the red wing of \CaIIK.

The \Heline\ emission is not confined to the red wing of the line, but appears as wide emission humps as illustrated in Fig.~\ref{fig:emission_examples}. The most extreme profiles exhibit emission out to large Doppler shifts from the line core, ranging from 180\,\kms\ blueshift to 140\,\kms\ redshift.

\subsection{Analysis of emission sites}

\begin{figure*}
\sidecaption
\includegraphics[width=12cm]{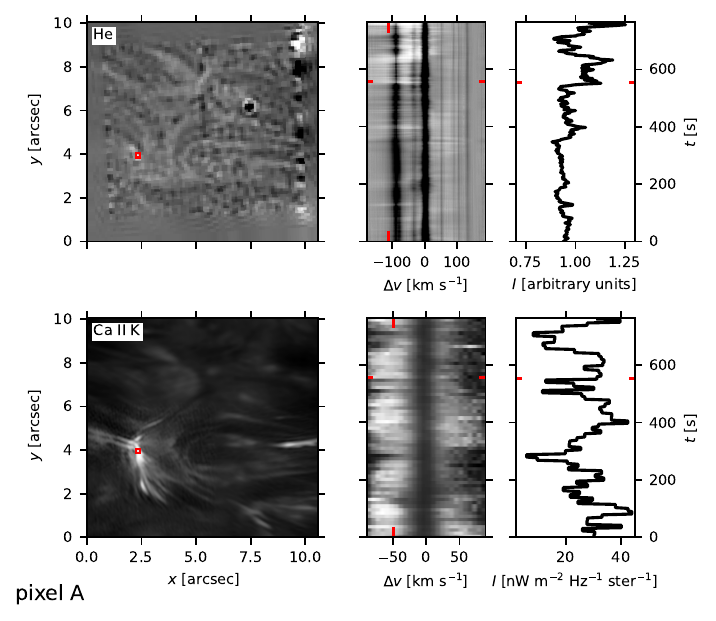}
\caption{Emission in pixel A. Upper left: image of the scaled intensity $I_\lambda(x,y,t)/I_{\mathrm{cont}}(x,y,t)$ in \Heline. Upper-middle: $\lambda t$-slice of the scaled intensity at the location of the red square in the image. The red tick marks indicate the wavelength and time of the image. Upper right: light curve at the spatial location indicated by the square and the wavelength indicated by the tick marks. The bottom row shows the same for \CaIIK, except that the image in the lower-left panel shows $I_\lambda$, i.e., it is not scaled to the local continuum intensity. }
\label{fig:lightcurve_A}
\end{figure*}

\begin{figure*}
\sidecaption
\includegraphics[width=12cm]{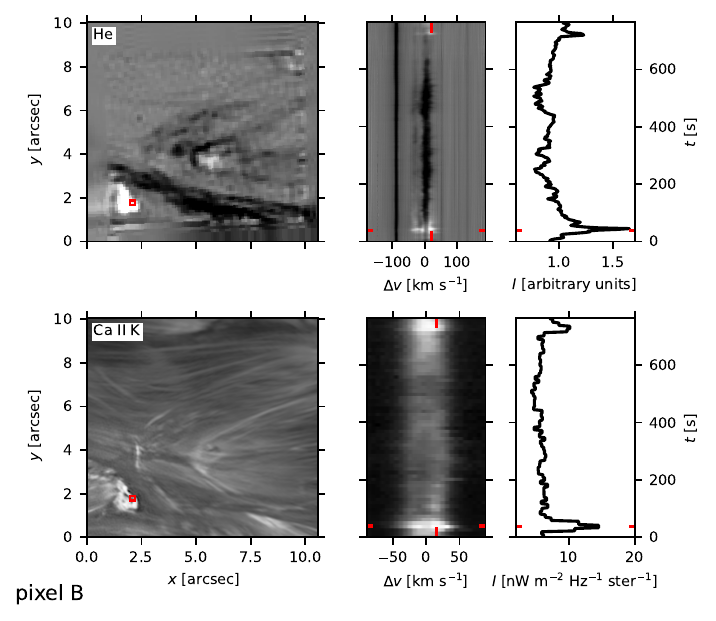}
\caption{Emission in pixel B. The figure format is the same as in Fig.~\ref{fig:lightcurve_A}.}
\label{fig:lightcurve_B}
\end{figure*}

\begin{figure*}
\sidecaption
\includegraphics[width=12cm]{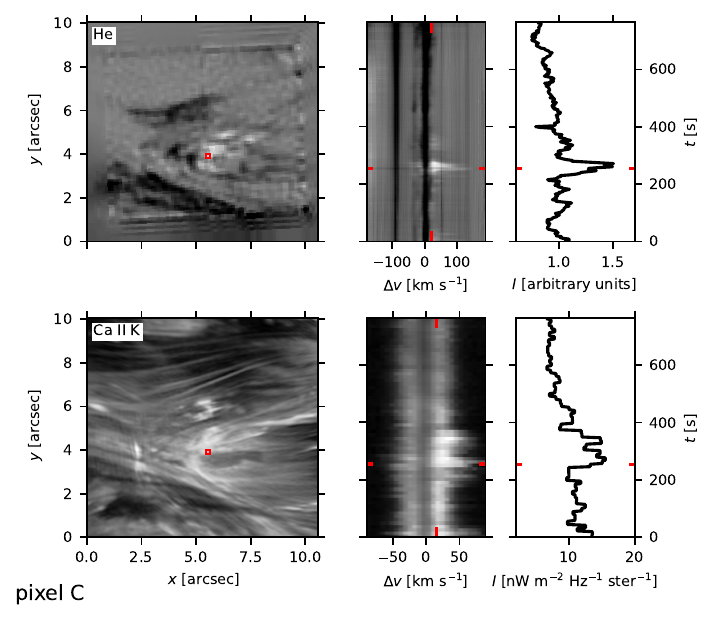}
\caption{Emission in pixel C. The figure format is the same as in Fig.~\ref{fig:lightcurve_A}.}
\label{fig:lightcurve_C}
\end{figure*}

In the following sections we present a deeper analysis of three representative pixels that show emission. These pixels are designated as follows: pixel A (red in Fig.~\ref{fig:emission_examples}), pixel B (brown), and pixel C (orange). Figures~\ref{fig:lightcurve_A}\,x--\,\ref{fig:lightcurve_C} show the locations of these pixels, their time evolution and a comparison with the \CaIIK\ line. Figure~\ref{fig:3panel_inversion} shows the fits obtained with the stacked-slab inversions as described in Sec.~\ref{sec:model_fits}, and the inferred parameters are given in Table~\ref{tab:fit_values}. We note that the optical depth and source function in the emission slab suffer from significant degeneracy. The emission peaks tend to be well-fitted both by a modestly high source function and an optically thick slab, as well as a very high source function with a slab thickness below unity. We show examples of these two types of fits in appendix~\ref{sec:appendix}.

\subsection{Pixel A} 

The \Heline\ line profile of Pixel A shows a very wide emission component that is located on top of the \Siline\ line, raising the question of whether the emission is really caused by \Heline. Because the \Siline\ line core is in absorption, the emission, were it caused by \ion{Si}{I}, would have to be caused by a high source function deeper down than the \Siline\ line core formation height of $\approx 500$~km 
\citepads{2008ApJ...682.1376B}. 
However, comparison with the movie of Fig.~\ref{fig:all_images} shows that the emission is not visible in the \Feline\ line and we conclude that the emission is indeed caused by \Heline.

Panel~{\it e} of Fig.~\ref{fig:all_images} and the corresponding movie show that the emission is located at a site of persistent photospheric magnetic flux cancellation. This cancellation causes persistent emission in \CaIIK\ (panel~{\it l}) and \Halpha\ (panel~{\it p}) with an elongated shape. The line-core images (panels {\it i)} and {\it k)} show a faint hint of the emission peeping through the fibril canopy, but is invisible in the \Halpha\ core (panel~{\it o)}.
The persistence of the \CaIIK\ emission is also clearly seen in the lower panels of Fig.~\ref{fig:lightcurve_A}. Taken together, we obtain a picture of reconnection brightening in the chromosphere, but clearly below the fibril canopy, reminiscent of Ellerman bombs.

In contrast, the \Heline\ emission is much more roundish and blob-like, and more intermittent in time as seen in the upper panels of Fig.~\ref{fig:lightcurve_A}. As can be seen by comparing the rightmost panels of Fig.~\ref{fig:lightcurve_B}, the \CaIIK\ brightness at times of the  \Heline\ emission does not stand out. 

The line profile is well-fitted by the Hazel inversion (Fig.~\ref{fig:3panel_inversion}). The upper slab that represents the fibril canopy is marginally optically thick ($\Delta \tau = 2$, see Table~\ref{tab:fit_values}), essentially at rest and with a LOS magnetic field strength that is below our noise limit. The layer that is responsible for the blueshifted emission has an upflow speed of 79~\kms. The Doppler width has an unrealistically high value of 39~\kms\ corresponding to a temperature of $\sim400$~kK. This is too high for the existence of significant amounts of \ion{He}{I}, even when accounting for non-equilibrium ionisation
\citep{2014ApJ...793...87Z}.
The Doppler width should thus be interpreted as the presence of large velocity gradients along the line-of-sight and/or within a HeSP resolution element that cannot be fitted with a constant-property slab
\citep[see also][]{2007A&A...462.1147L}.
Likewise, the fitted value of the the Voigt damping parameter $a=1.0$ is unphysically large and must be interpreted as the inversion trying to fit unresolved velocity gradients, because even at photospheric densities the damping parameter is of order $\approx 0.01$.

\subsection{Pixel B} 

Pixel B is part of a slightly larger brightening event with about one arcsec diameter located just outside of the fibril canopy, at $(x,y)=(2\arcsec,2\arcsec)$ in Figs.~\ref{fig:all_images} and ~\ref{fig:lightcurve_B}. It is located close to, but not on top of, the photospheric polarity inversion line, and is therefore probably not an Ellerman-bomb-like reconnection event. The emission in \Heline\ is as concentrated in time as in \CaIIK. It is very bright for only 10~s after which the emission rapidly weakens and fades away completely after 30 s. 
  
The Hazel inversion (Fig.~\ref{fig:3panel_inversion}) properly fits the Stokes $I$ and $V$ profiles. Because the emission is outside the fibril canopy, the properties of both the upper and the lower slab are likely set in the emission region. This is evidenced by the similar values of $v_{\mathrm{LOS}}$ and $\Delta{v_D}$. The source function in the upper slab ($\beta=1$) is lower than in the lower slab ($\beta=8$) in order to fit the central reversal of the line. We speculate that the observed central reversal is just a consequence of scattering line formation where the source function drops below the Planck function at heights where the line core forms (similar to the formation mechanism of the central reversal in the \CaIIHK\ and \MgIIhk\ lines).

Interestingly, the Stokes $V$ signal in this pixel is strong enough to fit a reasonable value of the LOS magnetic field, $-509$~G in the upper slab, and $-665$~G in the lower slab. Given the simplicity of the slab model, we do not believe that the inferred difference in field strengths is necessarily real, but it is clear that the emission forms in a layer with a magnetic field of about 0.5--0.6~kG. This is consistent with the value derived from applying the weak-field approximation to the \Caline\ at the same location.

\subsection{Pixel C} 

The brightening in Pixel C occurred at $(x,y)=(6\arcsec,4\arcsec)$ at $t=245$~s above the pore in the middle of the FOV of HeSP, several arcseconds away from a photospheric polarity inversion line. It is part of a cluster of blobby or slightly elongated brightenings of sub-arcsecond size that appear to move to the left as time passes, as can be seen in panel~{\it j)} in the movie of Fig.~\ref{fig:all_images}. A second \Heline\ emission event occurs in the same area around $t=403$~s. Inspection of the data (not shown here) reveals that most of the emission in this area shows either no clear Doppler shift, or shows redshift.

Comparison between the \Heline\ and \CaIIK\ images in Fig.~\ref{fig:lightcurve_C} shows that \CaIIK\ is bright at the same locations as \Heline, but it also shows bright fibrils that emanate from the location of the \Heline\ emission. The \CaIIK\ is more persistent in time as well, with persistently elevated brightness between $t=0$ and $t=400$~s.

The emission is neither visible in the line core of the photospheric \Feline\ (Fig.~\ref{fig:all_images}n) nor the cores of the chromospheric \Heline\ and \CaIIK\ lines, and the emission is thus most likely located in the 
mid-chromosphere, just as for Pixel~A.

\begin{figure*}[t]
\centering
\includegraphics[width=\textwidth]{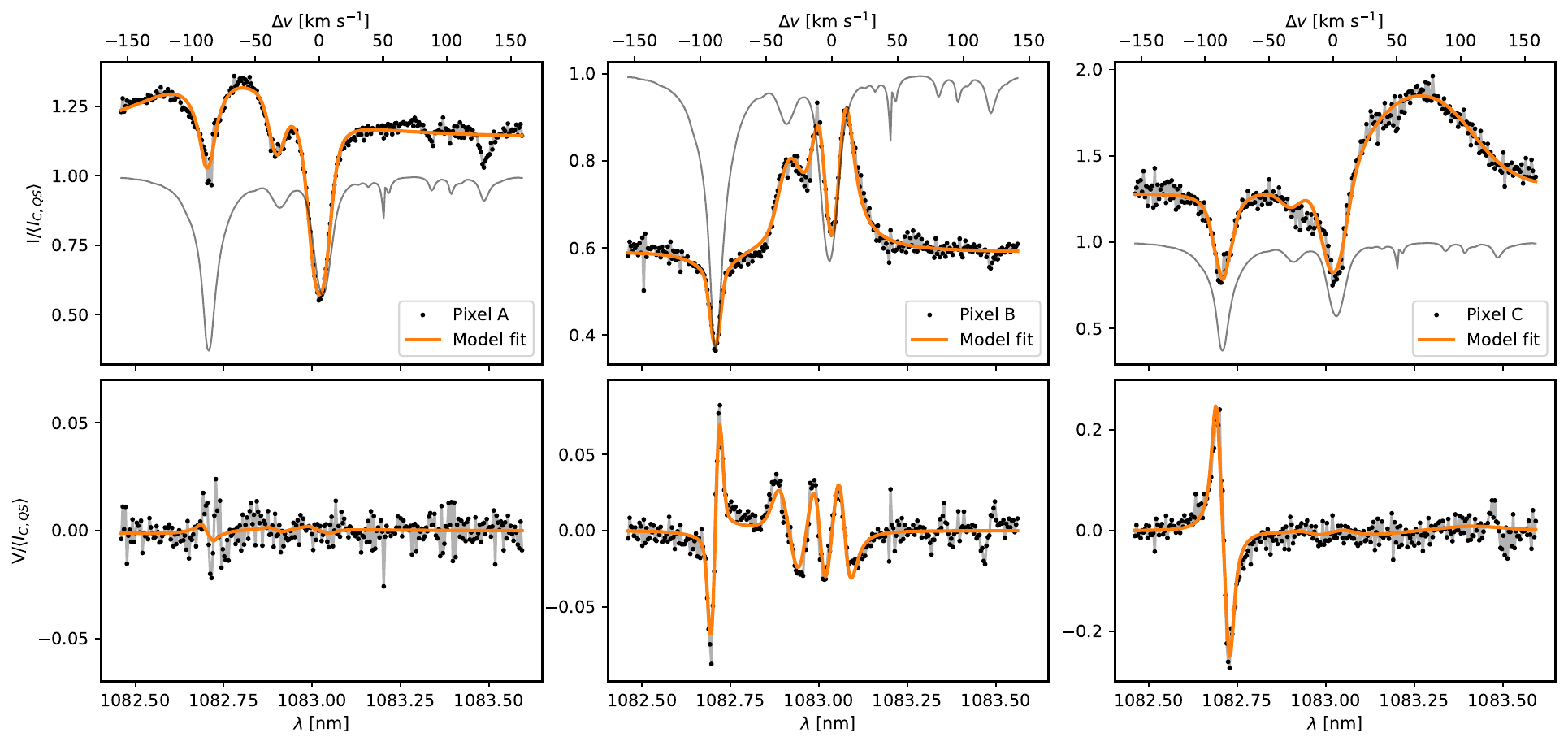}
\caption{Inversion results for the three selected pixels with emission in the \Heline\ line. The columns show the results for pixels A (left), B (middle), and C (right). The top row shows Stokes $I$; the bottom row Stokes $V$. Black-filled circles represent the observations, the orange solid lines are the best-fit profiles. The thin grey curve represents an average of a relatively quiet part of the FOV.}
\label{fig:3panel_inversion}
\end{figure*}

The inversion code fits the line profile well, and predicts an upper slab that is at rest, with a modest Doppler width. The lower slab, representing the emission site, exhibits a downflow of 67~\kms. Just as for Pixel A, the fitted microturbulence is unrealistically high at 32~\kms, indicating strong unresolved velocity gradients along the line of sight and/or within a HeSP resolution element. 
The inferred value of the LOS magnetic field in the lower slab is not reliable: the lower right panel of Fig.~\ref{fig:3panel_inversion} shows that Stokes $V$ in the emission hump is dominated by noise.

Because Pixel~C is located away from a photospheric polarity inversion line, we deem Ellerman-Bomb-like reconnection an unlikely cause of this brightening. The azimuth of the magnetic field in the chromosphere shows an elongated patch of NE-SW orientation (blueish in Fig.~\ref{fig:all_images}h) that makes a roughly $90^{\circ}$ angle with the predominant NW-SE orientation (red-brown in Fig.~\ref{fig:all_images}h). The \Heline\ brightening is located at the left end of this patch at $x=6$~Mm. The inclination of the chromospheric field (Fig.~\ref{fig:all_images}g) shows two patches of a close-to-vertical field on both sides of the patch with NE-SW azimuth. The movie of Fig.~\ref{fig:all_images} shows that the blue patch in azimuth starts to appear around $t=200$~s, and has disappeared by $t=250$~s.  

The chromospheric magnetic field was inferred with the weak field approximation, and lacks information on the geometrical height of the inferred magnetic field. This means we cannot determine the detailed magnetic field configuration around Pixel~C. Nevertheless, the large gradients in azimuth and inclination around Pixel~C mean that the field must exhibit shear, which in turn implies the presence of electric currents, and possibly also magnetic reconnection. 

\begin{table}\small
    \caption{Retrieved parameters and associated uncertainties using the stacked {\sc Hazel-2} slab model to the observed spectra. Values without uncertainties were kept fixed during the inversion.}
        \begin{tabular}{l r l r l r l}	
        \hline
        \hline	
        & \multicolumn{2}{c}{Pixel A} & \multicolumn{2}{c}{Pixel B} & \multicolumn{2}{c}{Pixel C} \\
        \hline
        \\
        \textbf{Upper slab} & & & & & & \\ 
        $v_{\rm LOS}$ [\kms] & -1.5 & $\pm 0.1$ & +2.8 & $\pm 0.2$ & -1.1 & $\pm 0.1$ \\ 
        $\Delta\tau$     & +2.0 & $\pm 0.3$ & +1.0 & $\pm 0.6$  & +1.1 & $\pm 0.5$ \\ 
        $\Delta v_D$ [\kms] & +5.6 & $\pm 0.1$ & +6.6 & $\pm 2.1$ & +8.4 & $\pm 0.1$  \\ 
        $B_{\rm LOS}$ [G]   & +17 & $\pm 15$  & -509 & $\pm 90$ & -80 & $\pm 17$ \\ 
        $\beta$        & +1.0  &  & +1.0  & & +1.0 & \\ 
        $a$            & +0.5  & & +0.5 && +0.5 &  \\  
    \\
        \textbf{Lower slab} & & & & & &  \\ 
        $v_{\rm LOS}$ [\kms] & -77.9 & $\pm 0.9$  & +3.2 & $\pm 0.1$  & +67.0 & $\pm 0.5$ \\ 
        $\Delta\tau$   & +1.0 & $\pm 0.7$ & +2.3 & $\pm 6.0$ & +2.0 & $\pm 1.2$ \\ 
        $\Delta v_D$ [\kms]  & +39.4 & $\pm 1.0$ & +7.1 & $\pm 3.0$ & +32.1 & $\pm 6.4$ \\ 
        $B_{\rm LOS}$ [G]    & +296 & $\pm 560$& -665 & $\pm 38$ & +636 & $\pm 840$ \\ 
        $\beta$          & +3.5& $\pm 2.8$  & +8.0 & $\pm 11$ & +4.0 & $\pm 5.0$ \\ 
        $a$               & +1.0 &  & +0.5 & & +0.5 &  \\
        $T_{\mathrm{exc}}$ [kK]       & 7.4   & & 11.4  & & 7.9& \\ 
            \hline
        \end{tabular}
    \label{tab:fit_values}
\end{table}

\section{Interpretation based on numerical modelling} \label{sec:modelling}

\begin{figure*}
\centering
\includegraphics[width=\textwidth]{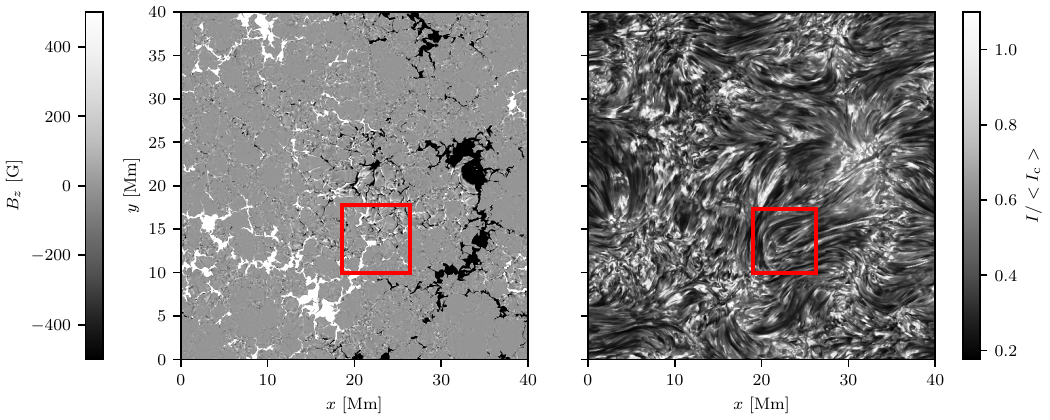}
\includegraphics[width=\textwidth]{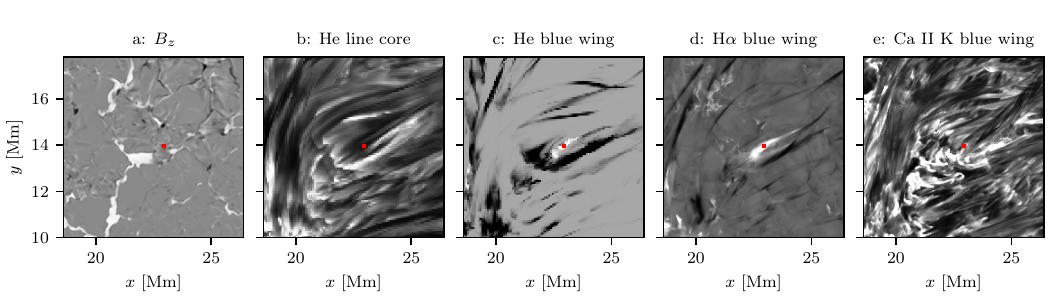}
\caption{Simulation data. Top row: Vertical component of the magnetic field at $z=0$~km and emergent intensity at the nominal line core of the strong component of the \Heline\ line. Bottom row: zoom in of the region inside the red square in the top row. Panel {\it a:} Vertical component of the photospheric magnetic field at $z=0$~km; panel {\it b:} emergent intensity at the nominal line core of the strong component of the \Heline.; panel {\it c:} local continuum scaled intensity of the \Heline\ at a Doppler shift of -18~\kms; panel {\it d:} \Halpha\ intensity at a Doppler shift of -33~\kms; panel {\it e:} \CaIIK\ intensity at a Doppler shift of -8~\kms. The pixel marked with the filled red square is analysed in Fig.~\ref{fig:fourpanel}.}
\label{fig:sim}
\end{figure*}

\begin{figure*}
\sidecaption
\includegraphics[width=12cm]{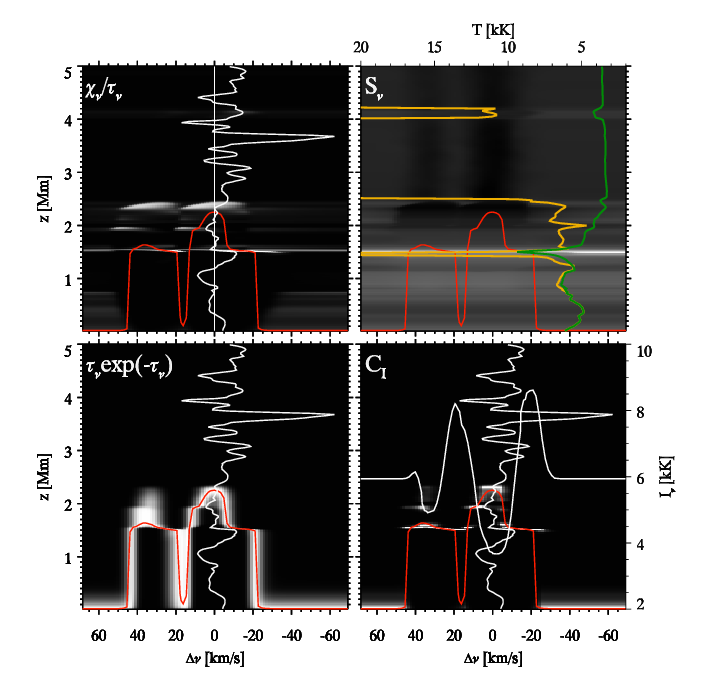}
\caption{Four-panel formation diagram of the pixel marked in red in the bottom row of Fig.~\ref{fig:sim}. {\it Top left:} Opacity $\chi_\nu$ divided by optical depth $\tau_\nu$. The red curve is the height optical depth is unity, the white vertical line marks $\Delta \nu = 0$ and the white curve is the vertical velocity as a function of height. 
{\it Top right:} total source function. The green curve shows the line source function, the orange curve the temperature.
{\it Bottom left:} $\tau_\nu \exp^{-\tau_\nu}$, which peaks at $\tau_\nu=1$.  
{\it Bottom right:} the contribution function $\chi_\nu S_\nu \exp^{-\tau_\nu}$. The white curve shows the emergent line profile in temperature units, with a scale on the right.}
\label{fig:fourpanel}
\end{figure*}

The formation of the \Heline\ line in an Ellerman Bomb was studied by
\citetads[][hereafter L21]{2021A&A...652A.146L}
based on a radiation-MHD simulation of flux emerging into the atmosphere by
\citetads{2019A&A...626A..33H}.
In this simulation, reconnection occurs in a long vertical current sheet in the chromosphere as a consequence of the squeezing together of opposite polarity flux in the photosphere.  As a result, a sheet of material at coronal temperatures is formed, as well as a bidirectional jet. 
Figure 5 in L21 shows that the \Heline\ line profiles at this current sheet exhibit broad emission peaks with intensity up to twice the continuum intensity, and with Doppler shifts ranging from -100~\kms\ to +80~\kms. Such profiles are reminiscent of the observed profiles presented in this work. 

L21 present an analysis of the formation of one of their synthetic line profiles that exhibits both a redshifted and a blueshifted emission component in their Fig.\,7, which shows a four-panel formation diagram following the format of 
\citetads{1997ApJ...481..500C}.

The red-shifted component forms in the downward part of the bidirectional jet in an emitting layer of about 100~km thickness. This layer is located in a strong velocity gradient which causes a large non-thermal width, similar to what we find in our Hazel inversions. The blue-shifted emission component forms at a larger height, in the upflowing part of the bidirectional jet. It is likewise formed in a thin layer with a gradient in the vertical velocity. Finally, the line profile shows a central depression formed at even larger heights, and is caused by the overlying fibril canopy (see Fig. 4 in L21).

Our Pixel~A is located above a patch of persistent photospheric flux cancellation and shows EB-like behaviour, similar to the situation described in L21.  The line profile in Pixel~A can be well-fitted with only one emission component, which is blueshifted. We therefore interpret this emission as formed in the upwardly-directed reconnection outflow. Whether the reconnection gives rise to a bidirectional flow as in L21 requires full non-LTE inversions as in
\citetads{2019A&A...627A.101V}, 
which is beyond the scope of the present paper.  

Pixel B is not located above a photospheric polarity inversion line. The magnetic field appears to be quite vertical in both the photosphere and chromosphere, even though there is a visible substructure in the inclination at the site of the brightening (see the movie of Fig.~\ref{fig:all_images} around 36~s). The azimuth of the field in the chromosphere shows a large gradient of up to 90 degrees across the brightening (panel~h in the movie). We could not infer a likely field configuration from the retrieved magnetic field information so we must refrain from interpreting this event based on the numerical modelling.

Pixel~C is part of a complex of brightenings located in a region away from the bipolar magnetic field in the photosphere. The magnetic configuration in the chromosphere exhibits strong changes in the azimuth of the field. The snapshot analysed by L21 did not contain such an event. Instead, we searched for and found one event in the MURaM simulation described in Sec.~\ref{sec:simulations} that shows a brightening event in the \Heline\ associated with a rapid change of field azimuth with height. 

The top row of Fig.~\ref{fig:sim} shows the entire simulation domain. There are two large elongated patches of strong photospheric field, and in between them, there is ongoing flux emergence (inside and above the red box). The \Heline\ line core shows a complex fibrillar structure that in general connects the patches of opposite polarity. The appearance of the fibrils in the line core is reminiscent of our observations (see Fig.~\ref{fig:all_images}i). 

In the bottom row of Fig.~\ref{fig:sim} we show the region with the \Heline\ brightening. Panel~c shows the local-continuum-divided \Heline\ intensity at a Doppler shift of $\Delta v = -18$~\kms,
and exhibits an elongated patch with intensity higher than the local continuum. 
The photospheric magnetic field in panel~{\it a} shows the brightening is associated with field lines emanating from a unipolar photospheric magnetic field concentration, without significant opposite polarity flux nearby. 

The brightening in the \Heline\ wing is not visible in the line core image in panel~b. The blue wing of \Halpha\ (panel~d) shows the same brightening, but the \CaIIK\ (panel~e) wing does not. Owing to the large difference in cadence between the \CaIIK, \Halpha, and \Heline\ data we cannot be sure of the observational configuration, but Fig.~\ref{fig:all_images} indicates the Pixel C event is visible in all three lines, in contrast to our simulation. 

Figure~\ref{fig:fourpanel} displays the line formation for the brightening through a four-panel formation diagram
\citep{1997ApJ...481..500C}
of the pixel indicated in the lower row of Fig.~\ref{fig:sim}. The emission peaks are formed in a single thin layer at $z=1.5$~Mm, where the line source function shows a sharp local maximum. This height is the $\tau=1$ height at the frequencies of the emission peaks in the line profile. The source function maximum is located right next to a thin layer with a maximum temperature of $T=29$~kK.
Determining whether the source function maximum is caused by the photoionisation-recombination mechanism or direct collisional ionisation/excitation would require the same extensive analysis as in L21, and because the effect on the line profile is the same, we refrain from doing that here. The emission layer shows a downflow with a speed of 12~\kms. The absorption throughs are formed in a range of heights between $z=1.6$~Mm and $z=2.4$~Mm, where the material has a vertical velocity in the range from $-5$~\kms\ to $+4$~\kms.

This analysis shows that also for this type of event, our simple inversion model of a hot emitting layer below an absorbing cooler fibril layer appears valid. While the absorbing layer has a vertical extent of 0.8~Mm and shows a range of velocities, the source function in this layer is rather constant, and the inversion would likely be able to fit the profile with a single source function and some microturbulence. 
 The downflow in the emitting layer, combined with the roughly zero velocity of the absorbing layer causes the leftmost emission peak to be lower than the other two (as seen of the relatively low value of $\chi_\nu/\tau_\nu$ in the leftmost peak in the upper left panel of the figure.

\begin{figure*}
\sidecaption
\includegraphics[width=12cm]{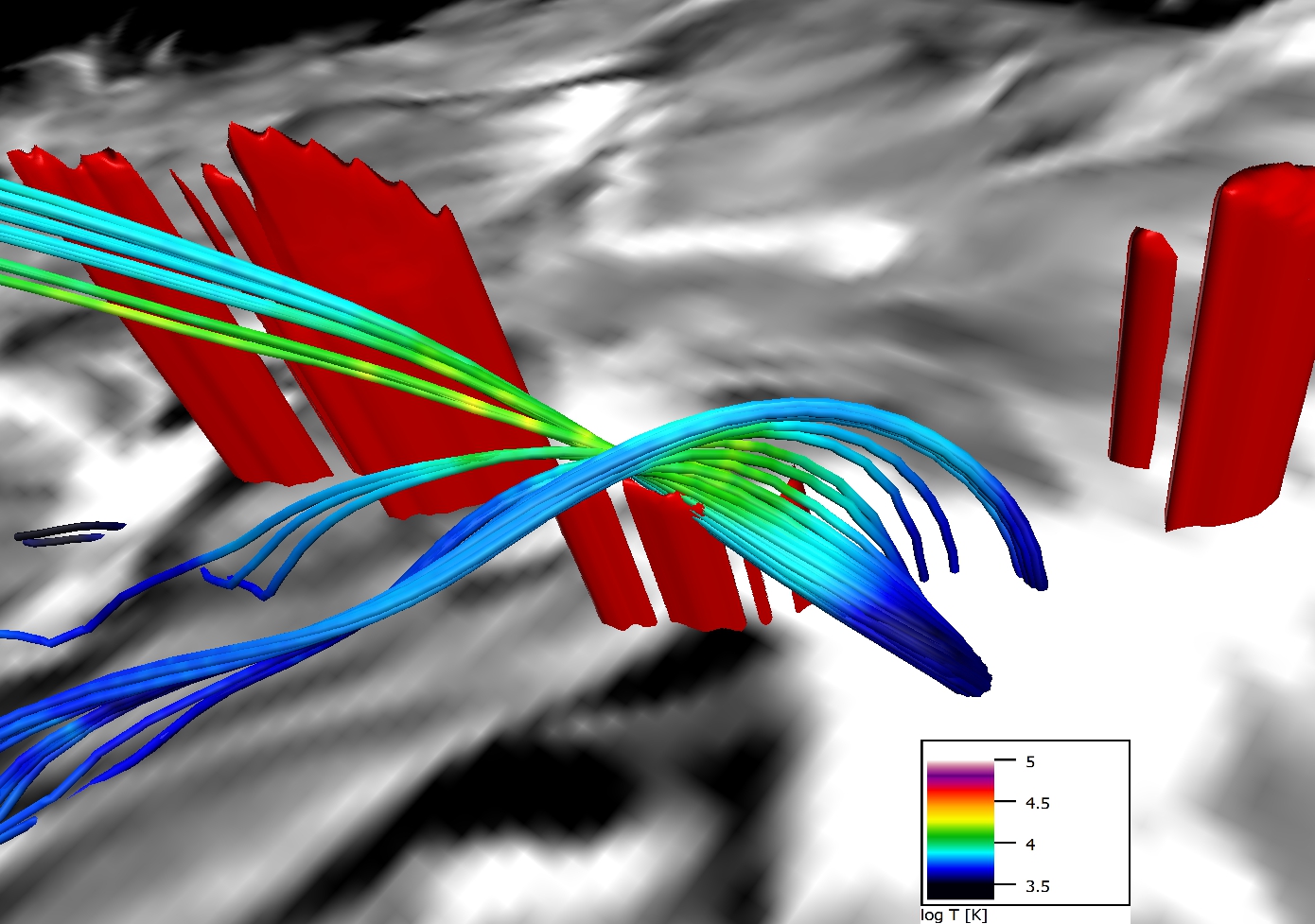}
\caption{3D visualisation of the magnetic field configuration of the \Heline\ emission event in our MURaM simulation. The greyscale background shows the vertical magnetic field in the photosphere (white: positive; black: negative), just above the continuum formation height. 
The red surface displays the height of optical depth unity $z(\tau=1)$ for the vertically emergent intensity at $\Delta v = -18$~\kms for the \Heline\ line, corresponding to the blue wing images in Fig.~\ref{fig:sim}. This height tends to lie below the height for which we display the magnetic field, and is above it only in some locations, appearing as steeply-sided columns.
The magnetic field lines thread the emitting region of the \Heline\ brightening, and their colour indicates the local temperature.
 Figure made with VAPOR
\citep{atmos10090488}.
}
\label{fig:vapor}
\end{figure*}

In Fig.~\ref{fig:vapor}, we analyse the magnetic field configuration of the event. The material causing the emission peaks is located at the top of the red pillars in the middle of the image, which represent the $z(\tau=1)$ at $\Delta v = -18$~\kms\ for the \Heline\ line. The magnetic field lines that thread this region are strongly sheared, and a current sheet (not shown in the figure) is located next to the emission region. This current sheet causes the high temperature at $z=1.4$~Mm in Fig.~\ref{fig:fourpanel}.  The sheet is so thin that even though some of the field lines intersect with the sheet, this high temperature is not visible in the figure. 

The field lines originate from a unipolar region to the right of the current sheet. Even though there is opposite polarity flux right below the current sheet, this flux is unrelated to the sheet. 

Based on the analysis of this event in our simulation we conclude that the observed behaviour of pixel~C is consistent with current sheets in the chromosphere caused by sheared magnetic field lines emanating from unipolar regions in the photosphere.
\citep[see also][]{2015ApJ...811..106H,2023A&A...672A..47S}
The Joule heating inside the sheet can be sufficiently large to raise the temperature to at least 20~kK, sufficient for either the photoionisation/recombination mechanism or direct collisional excitation to create \Heline\ opacity and a line source function larger than the source function in the photosphere.

\section{Discussion \& Conclusions}\label{sec:discussion}
 
In summary, we find small fast-evolving brightenings in the \Heline\ line in the observations. They are often cospatial with \Halpha\ and \CaIIK\ brightenings, but the latter are much more persistent in time and extended in space. 
As \Heline\ only shows emission in or next to sites with temperatures above 20~kK, these emission sites must indicate small hot spots in the lower solar atmosphere, below the fibril canopy. 
 
A two-layer inversion model can fit the observed \Heline\ profiles, and we find large upflows and downflows of order $\pm 75$~\kms. These speeds are inconsistent with acoustic phenomena (the sound speed is of order 10~\kms) as well as observed velocity amplitudes in transverse or torsional waves in spicules
\citep{2012ApJ...752L..12D}
 and fibrils
\citep[e.g.][]{2012NatCo...3.1315M,2022arXiv221014089K}.

These flow speeds are instead consistent with the Alfv{\'e}n speed $\va$. Taking $B = 665$~G as found for the lower slab in pixel~B, and a mass density $\rho = 7 \times 10^{-6}$~kg~m$^{-3}$, corresponding to $z=500$~km in the FALC model atmosphere
\citep{1993ApJ...406..319F},
we find $\va = 22$~\kms. For $\rho = 10^{-7}$~kg~m$^{-3}$, corresponding to $z=1000$~km in FALC, $\va = 188$~\kms.

Reconnection outflow speeds tend to be on the order of the Alfv{\'e}n speed
\citep[e.g.,][]{1957JGR....62..509P,2014masu.book.....P},
and we interpret the speeds that we recover for pixels~A and~C as caused by such outflows. To our knowledge, this is the first time such line-of-sight flow speeds are measured in reconnection events at sub-arcsecond scales. Earlier inversions of Ellerman bombs and UV bursts based on \CaIIK, \MgIIhk\ yielded velocities up to $\pm 30$~\kms 
\citep{2019A&A...627A.101V}.
Inversions of Ellerman Bombs in the \HeI~D$_3$ and \Heline\ lines by 
\citet{2017ApJ...851L...6R}
found velocities from 22~\kms\ downflows up to 40~\kms\ upflows. 
\citet{2021A&A...647A.188D} found 
plasmoids visible in the \CaIIK\ line at a blueshift of 100~\kms, but did not determine the flow speed with inversions. 

Speeds of order 100~\kms\ derived from the \Heline\ line have been measured before
\citep[e.g.][]{2011A&A...526A..42S,2019A&A...625A.128D,2022A&A...661A.122S},
but with absorption components and for filaments and along chromospheric fibrils at larger spatial scales than reported here.

UV bursts seen in the  \SiIV~139.4~nm and~140.3~nm lines
\citep{2014Sci...346C.315P},
are a signature of low-atmospheric reconnection. At least a fraction of them is caused by the same mechanism that produces Ellerman Bombs: the squeezing together of opposite polarity fields in the photosphere and chromosphere
\citep{2019A&A...626A..33H},
 and flow speeds up to $\pm 100$~\kms\ have been reported for them
\citep{2014Sci...346C.315P,2018ApJ...854..174T}.
UV bursts tend to have an observed size of order $1\arcsec$\,--\,$3\arcsec$, substantially larger than our pixels A and~C, but comparable to pixel~B. 

Magnetic reconnection in three dimensions is extremely complex and can occur in a variety of magnetic field configurations
\citep{2022LRSP...19....1P}.
We showed using 3D radiation-MHD simulations that some of the observed \Heline\ profiles are consistent with opposite polarity reconnection of vertically oriented field 
\citep[pixel A, see][]{2021A&A...652A.146L}
%
and current sheets in sheared largely horizontal field in the chromosphere (pixel C, simulation presented in this paper). The limited spatial resolution and concomitant large numerical resistivity and viscosity in these simulations mean that the reconnection current sheets lack significant substructure. The resulting \Heline\ profiles  
\citep[Fig.~\ref{fig:fourpanel} and ][]{2021A&A...652A.146L}
therefore do not exhibit the wide emission humps (requiring large non-thermal broadening to fit) seen in our observations (Fig.~\ref{fig:emission_examples}).

Long thin current sheets are however unstable and fragment into plasmoids 
\citep[magnetic islands in 2D and twisted flux ropes in 3D, see Sec.~8 of][]{2022LRSP...19....1P}.
This process has only been studied in 3D under idealised or coronal circumstances
\citep[e.g.,][]{2016ApJ...818...20H,2023A&A...678A.132S}. 
In 2D it has been studied under chromospheric conditions in self-consistent 2D radiative-MHD simulations by 
\citet{2017ApJ...851L...6R}.
The exact spatial resolution of this simulation is not reported, but while the resolution was high enough to lead to fragmentation of the current sheet, the resulting plasmoids do not show internal substructure.

In the 2D simulations of 
\citet{2021A&A...646A..88N}
and 
\citet{2024A&A...685A...2C}
adaptive mesh refinement was used to simulate plasmoids in the chromosphere with a spatial resolution less than one kilometer. They find plasmoids with a diameter of about 100 km, typically consisting of a hot ($ \sim 100$~kK) and tenuous partial ring surrounding a dense core with a temperature less than 10~kK. The hot ring should produce copious amount of UV radiation that populates the \HeI\ triplet system through photoionisation/recombination in the core as well as the surrounding undisturbed chromosphere 
\citep{2016A&A...594A.104L}.
They could thus well explain the small roundish emission features close to the diffraction limit that we find in our data. 

In addition, the simulated plasmoids harbour substantial internal velocity structure, with flow speeds reaching up to about 20~\kms. Because our observations have a spatial resolution of 200~km, we would not resolve the internal velocity gradients in the plane-of-the-sky direction. 
The vertical extent of the contribution function makes it impossible to resolve such a structure along the line-of-sight. These two effects combined offer a natural explanation for the wide emission humps and non-thermal velocities of order 40~\kms\ that we find in our inversions.

This study is an initial exploration of the first HeSP dataset.  The reconnection brightenings are small, evolve fast and show structure at large Dopplershift from the nominal line center. Based on the temperature needed to produce them, we conclude that transition-region-like temperatures in the deeper layers of the active region chromosphere are more common than previously thought. 

It highlights the power of high-cadence integral-field spectropolarimetry combined with image reconstruction to reach diffraction-limited spatial resolution. Neither slit spectrographs nor wavelength-scanning imaging instruments are able to fully characterise these events. The \Heline\ is a unique diagnostic to study small-scale reconnection and heating in the chromosphere through its sensitivity to temperature above 20~kK, where \HI, \CaII, and \MgII\ are already ionised away. 

The study also shows the importance of multiple diagnostics to constrain the physical interpretation. Work is underway to further analyse the \Heline\ brightenings through a more in-depth investigation of their motion, more extensive inversions with Hazel and full non-LTE inversions of the \CaIIK, \Feline, and \Caline\ lines to better constrain the magnetic field configuration in these events. It would also be of interest to compute the \Heline\ profiles emanating from models of chromospheric plasmoids.


\begin{acknowledgements}
The Swedish 1-m Solar Telescope is operated on the island of La Palma by the Institute for Solar Physics of Stockholm University in the Spanish Observatorio del Roque de los Muchachos of the Instituto de Astrof{\'i}sica de Canarias. The Institute for Solar Physics is supported by a grant for research infrastructures of national importance from the Swedish Research Council (registration number 2021-00169). 
JL acknowledges financial support from the Knut och Alice Wallenberg Stiftelse (project number 2016.0019).
This project received funding from the European Union (ERC, MAGHEAT, 101088184 and WINSUN, 101097844). Views and opinions expressed are however those of the author(s) only and do not necessarily reflect those of the European Union or the European Research Council. Neither the European Union nor the granting authority can be held responsible for them.
This project has received funding from Swedish Research Council (2021-05613) and Swedish National Space Agency (2021-00116).
This research is supported by the Research Council of Norway, project number 325491 and through its Centers of Excellence scheme, project number 262622.
The radiative transfer computations were performed on resources provided by the National Academic Infrastructure for Supercomputing in Sweden (projects NAISS 2023/1-15 and NAISS 2024/1-14) at the PDC Centre for High Performance Computing (PDC-HPC) at the Royal Institute of Technology in Stockholm.
\end{acknowledgements}

\bibliographystyle{aa}
\bibliography{references}

\appendix
\section{Degeneracy between the scaling factor $\beta$ and the optical depth $\Delta \tau$} \label{sec:appendix}

\begin{figure*}
\centering
\includegraphics[width=\textwidth]{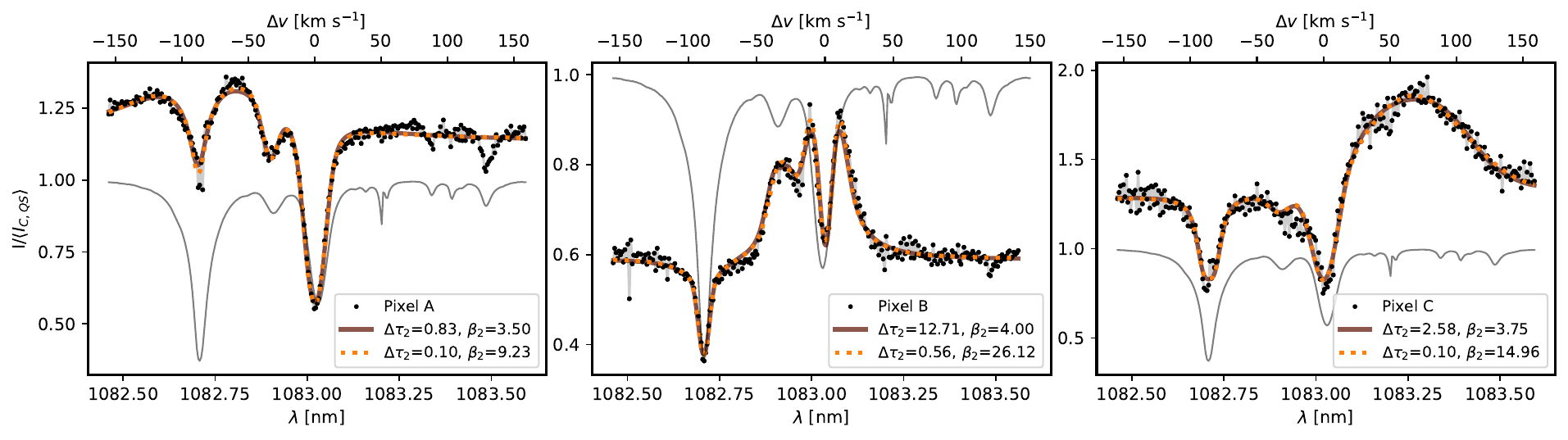}
\caption{Degeneracy between the scaling factor $\beta$ and the optical depth $\Delta\tau$. The figure shows the intensity profiles for the three selected pixels, with two models for each pixel, one with low $\beta$ and high $\Delta\tau$, and one with high $\beta$ and low $\Delta\tau$. The grey curve is a nearby quiet Sun spectrum.}
\label{fig:3panel_degeneracy}
\end{figure*}

We ran {\sc Hazel-2} inversions for the three selected pixels in Table~\ref{tab:fit_values} changing the value of $\beta$ to explore the degeneracy between $\beta$ and $\Delta\tau$ and quantify the range of possible values. The results are shown in Fig.~\ref{fig:3panel_degeneracy}. The figure shows the intensity profiles for the three selected pixels, with two models for each pixel, one with very low $\beta$ and high $\Delta\tau$, and one with high $\beta$ and low $\Delta\tau$. The figure shows that the two models are almost indistinguishable. From this experiment, we can conclude that we can modify $\beta$ from 3 to 20 and still obtain a good fit to the data. The rest of the parameters, in particular the magnetic field, are not affected by the choice of $\beta$.

\end{document}